# Hedgehog-like spin texture in Sb-doped MnBi$_2$Te$_4$

Meng Zeng[1]*, Shu Mo[1]*, Ke Zhang[2], Yu-Jie Hao[1], Yu-Peng Zhu[1], Xiang-Rui Liu[1], Cheng Zhang[3], Ming-Yuan Zhu[1], Shiv Kumar[3], Takuma Iwata[4,5], Koji Miyamoto[3], Taichi Okuda[3,6], Kenya Shimada[3,6], Kenta Kuroda[4,5,6], Xiao-Ming Ma[7,8]†, Chang Liu[1]†

[1] *Department of Physics, Southern University of Science and Technology (SUSTech), Shenzhen, Guangdong 518055, China*

[2] *School of Physics, University of Electronic Science and Technology of China, Chengdu, Sichuan 611731, China.*

[3] *Hiroshima Synchrotron Radiation Center, Hiroshima University, Higashi-Hiroshima, Hiroshima 739-0046, Japan*

[4] *Graduate School of Advanced Science and Engineering, Hiroshima University, Higashi-hiroshima, Hiroshima 739-8526, Japan*

[5] *International Institute for Sustainability with Knotted Chiral Meta Matter (WPI-SKCM$^2$), Hiroshima University, Higashi-hiroshima, Hiroshima 739-8526, Japan*

[6] *Research Institute for Semiconductor Engineering, Hiroshima University, Higashi-Hiroshima, Hiroshima 739-8527, Japan*

[7] *Shenzhen Institute for Quantum Science and Engineering, Southern University of Science and Technology (SUSTech), Shenzhen, Guangdong 518055, China*

[8] *International Quantum Academy, Shenzhen, Guangdong 518048, China*

## Abstract

We employ spin- and angle-resolved photoemission spectroscopy and circular-dichroism ARPES to systematically investigate the spin texture of Sb-doped MnBi$_2$Te$_4$. Our results display a hedgehog-like spin texture in this system which is signified by reversed-orienting out-of-plane spins at the Dirac gap. Our finding reveals the presence of time-reversal symmetry breaking, implying the possibility for realization of high-temperature quantum anomalous Hall effect.

* These authors contribute equally to this work.
† Corresponding Authors. E-mails: maxm3@sustech.edu.cn, liuc@sustech.edu.cn

The interplay of topology and magnetism has led to many novel physical phenomena, such as the quantum anomalous Hall effect (QAHE)[1–3], the axion insulator state[4–6], the spin Hall effect[7], and the magnetic topological superconductivity[8]. These effects hold great potential for future applications in topological quantum computing. Breaking the time-reversal symmetry and tuning the Fermi level into a magnetic band gap is crucial for realizing QAHE. The former ensures that only one chiral edge state participates in conduction[3], while the latter is key to achieving quantized Hall conductance[2,9]. The realization temperature of QAHE is closely related to the size of the magnetic gap. Enlarging the magnetic gap can suppress the destruction of QAH state caused by thermal excitation, enhancing the topological state's stability, and thereby effectively raising the QAHE temperature. The recently-discovered magnetic topological insulator $MnBi_2Te_4$ is an excellent platform for studying QAHE and the axion insulator state. The spontaneous magnetic order in $MnBi_2Te_4$ breaks time-reversal symmetry, theoretically opening a magnetic gap of the order of 100 meV in the topological surface state (TSS), which provides the possibility of achieving high-temperature QAHE[10]. However, experiments found that the actual gap size of $MnBi_2Te_4$ is much smaller than theoretically predicted[11–14], and the realization temperature of QAHE is also not as high as expected[15], which poses challenges for future applications.

Recently, Ma *et al.* reported that Sb substitution on the Bi site can simultaneously tune the Fermi level and the TSS gap size in $MnBi_2Te_4$[16]. In this system, the bulk energy gap and the magnetic gap of TSS are two key energy scales that are related to the realization of QAHE. A larger bulk energy gap is more favorable for band inversion to generate TSS, while a larger TSS gap can ease the tuning of the Fermi level into the gap and enhance the resistance of the QAH state to thermal fluctuations. Determining the nature of the gap is also important: if it is a trivial gap not induced by magnetism, QAHE cannot be realized even if the Fermi level is tuned into the gap[17]. Measuring the spin texture of the gapped TSS could pinpoint whether the gap is induced by the magnetic breaking of time-reversal symmetry. If the gap is induced by magnetism, its spin texture will be hedgehog-like, signified by large and sign-reversed out-of-plane spins at the two sides of the Dirac gap. If the gap is induced by chemical disorder on the surface, its spin texture will be helical-like, and no out-of-plane spin is present near the gap[18]. Although many exciting properties of Sb-doped $MnBi_2Te_4$ were suggested[19–25], the spin configuration of it remains unknown. Investigating the spin texture of Sb-doped $MnBi_2Te_4$ is not only essential for understanding the nature of its gapped TSS but also provides a reference for future applications in spintronics. In this study, we systematically investigate the spin texture of Sb-doped $MnBi_2Te_4$ by spin- and angle-resolved photoemission spectroscopy (SARPES) and circular-dichroism ARPES (CD-ARPES). We observe a hedgehog-like spin configuration in this system, which indicates that the gap is induced by the breaking of time-reversal symmetry.

In the doping range we studied, the symmetry of Sb-doped $MnBi_2Te_4$ is found to be the same as that in undoped $MnBi_2Te_4$, i.e., a rhombohedral crystal structure with a centrosymmetric space group $R\bar{3}m$. In this quasi-2D structure, seven atomic layers are arranged in a Te-X-Te-Mn-Te-X-Te (X = Bi/Sb) sequence, known as a septuple layer (SL). A van der Waals gap exists between adjacent SLs, as shown in Fig. 1(a). Fig. 1(b) presents the 3D and surface-projected Brillouin zone (BZ) of $Mn(Bi_{1-x}Sb_x)_2Te_4$, where $\bar{\Gamma}$-$\bar{M}$-and $\bar{\Gamma}$-$\bar{K}$ represent two high-symmetry directions of the 2D hexagonal BZ. To identify the high-symmetry directions of the samples, we employed Laue X-ray diffraction, as shown in Fig. 1(c). The elemental maps of all atoms in $Mn(Bi_{0.9}Sb_{0.1})_2Te_4$ are depicted in Fig. 1(d), revealing a homogeneous distribution of the elements. Fig. 1(e) shows the XRD patterns of Mn(Bi$_{1-x}$Sb$_x$)$_2$Te$_4$ for various nominal doping levels $x$. The minimal shifts in XRD peak positions with varying doping levels are consistent with previous report[20]. In this work, we estimate the actual doping levels of the samples based on the sizes of the TSS gap and the energy positions of the conduction band minimum, the same as in our previous work[16].

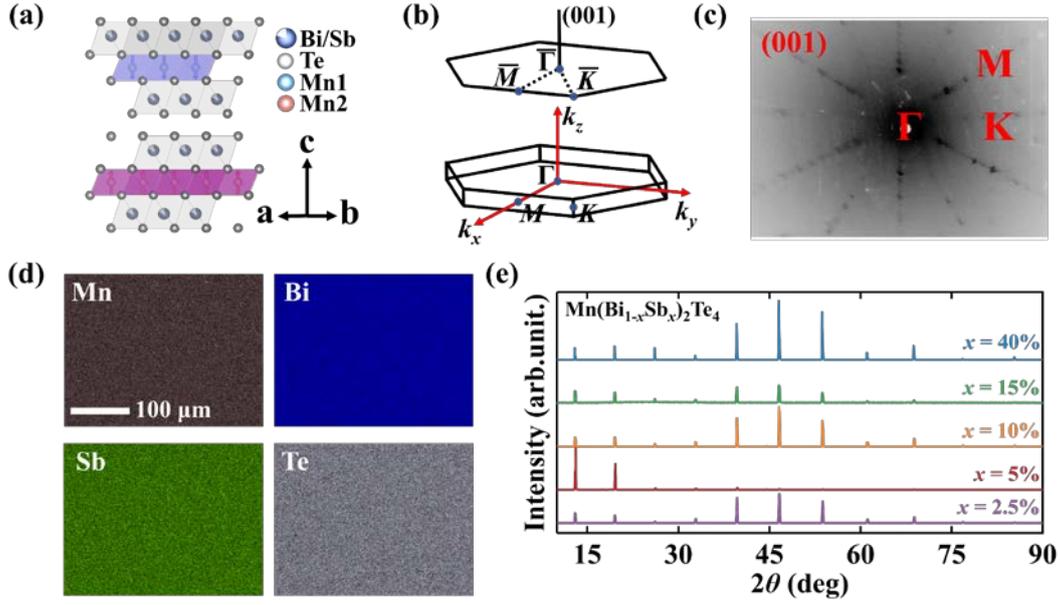

**Fig. 1. Basic properties of Sb-doped MnBi$_2$Te$_4$.** (a) Crystal structure of Sb-doped MnBi$_2$Te$_4$. The unit cell consists of two septuple layers with different magnetic moments indicated by the arrows. (b) The 3D and surface-projected BZ of Mn(Bi$_{1-x}$Sb$_x$)$_2$Te$_4$. (c) Laue X-ray diffraction pattern of the (001) surface of Mn(Bi$_{0.975}$Sb$_{0.025}$)$_2$Te$_4$. (d) Energy-dispersive X-ray spectroscopy (EDX) elemental maps of Mn(Bi$_{0.9}$Sb$_{0.1}$)$_2$Te$_4$. (e) Single crystal X-ray diffraction data for the samples on which our data was collected.

We use high-resolution laser SARPES with *p*-polarized light (light polarization parallel to the detection plane[26]) to investigate the spin texture of Sb-doped MnBi$_2$Te$_4$ with different doping levels. The measurements are conducted at 10 K (below the Néel temperature $T_N$ = 24 K), with the momentum direction parallel to $\bar{\Gamma}$-$\bar{M}$. Fig. 2(a) presents the spin-integrated ARPES results. The TSS gap gradually widens, and the conduction band minimum rises, as the doping level increases, reproducing the results in Ref. 16. Figs. 2(b)-(d) show the SARPES results along different spin directions. Fig. 2(e) presents the schematic diagrams of the ARPES-measured spin texture for both the conduction and the valence bands.

From our data, the spin textures for samples with different doping concentrations have commonalities and sample-dependent characteristics. On the one hand, for the conduction bands, the in-plane spin is approximately tangent to the momentum in all five samples measured, as the $S_y$ component always points out of the page at $k < 0$ and into the page at $k > 0$ [Fig. 2(c)]. Such spin-momentum locking exhibits a helical in-plane spin texture. The radial spin component $S_x$ also exists, which is smaller and has in general an opposite sign between the valence band and the conduction band at the same $k$ [Fig. 2(b)]. It has been suggested that radial spin polarization that is antisymmetric with respect to the zone center can be induced by the laser light in topological insulators such as $Bi_2Se_3$[27]. In this case, if the *k*-space location of the ARPES measurement is slightly off from $k_y = 0$, finite radial in-plane spin polarization can appear on both sides of $\bar{\Gamma}$, which may explain the observed $S_x$ pattern. Importantly, a substantial out-of-plane spin component ($S_z$) is observed for both the conduction and the valence band [Fig. 2(d)]. Under the laser light, $S_z$ is found to be always positive near the TSS gap, which seemingly differs from the spin texture of magnetically-doped $Bi_2Se_3$ and pristine $MnBi_2Te_4$ whose $S_z$ changes sign across the topological gap[16,18,28]. Although we primarily attribute such all-up configuration of $S_z$ to final-state effects (discussed below), we emphasize that the very existence of out-of-plane spin polarization on both sides of the gap is evident for a breaking of time-reversal symmetry, suggesting a magnetic origin of the TSS gap.

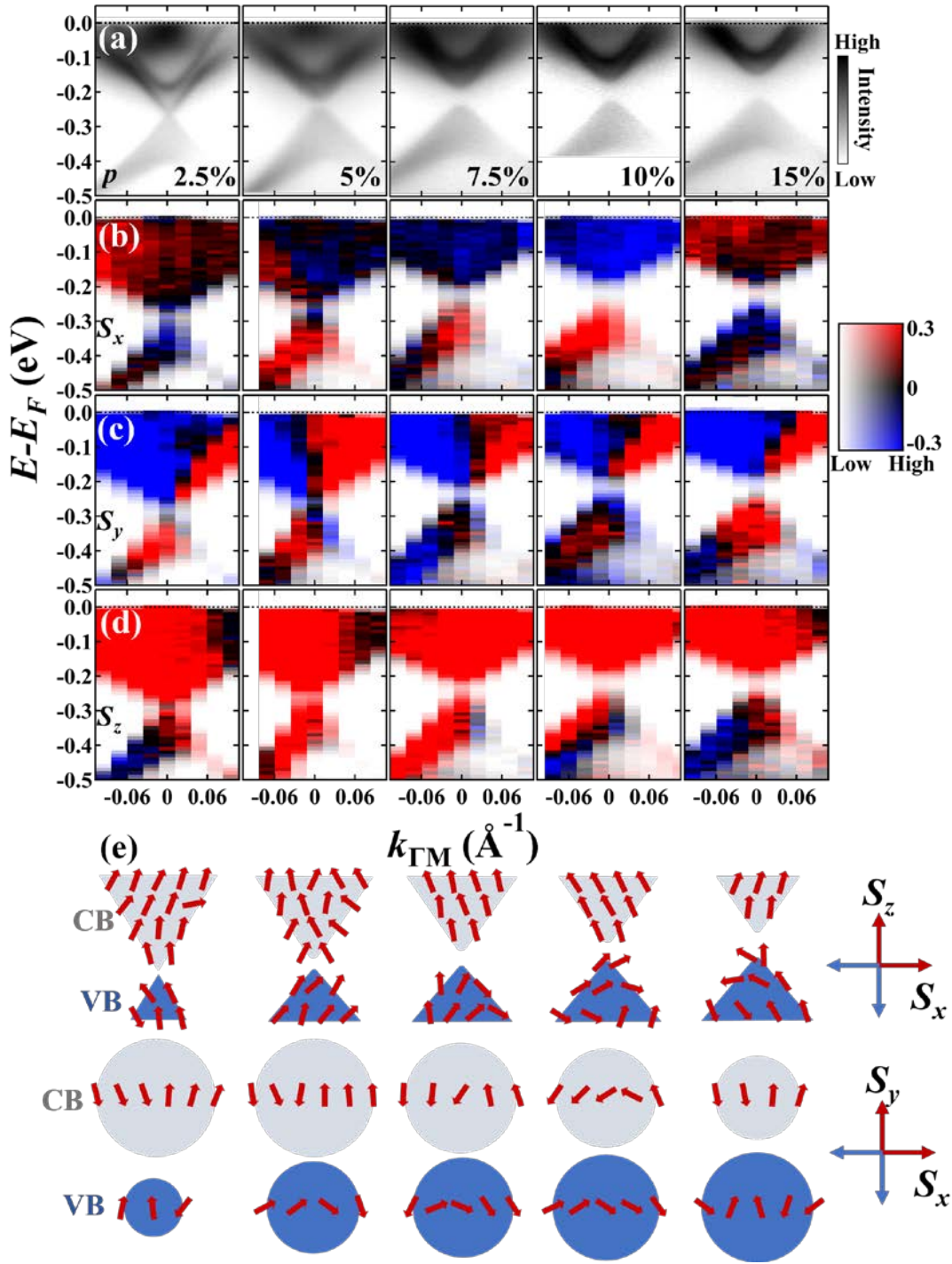

**Fig. 2. Spin-integrated and spin-resolved ARPES data for Mn(Bi$_{1-x}$Sb$_x$)$_2$Te$_4$ with different $x$, acquired at $h\nu$ = 6.3 eV.** Data was measured using $p$-polarized laser radiation at a temperature of 10 K (below $T_N$). (a) Spin-integrated ARPES data. $x$ values are labeled at the lower right. (b-d) Spin-resolved ARPES maps along $\bar{\Gamma}$-$\bar{M}$ for three spin directions, where the red and blue colors are defined at the right legend. (e) Schematic diagrams of the measured spin texture. The grey and blue colors represent the conduction bands (CB) and the valence bands (VB), respectively. Directions of red arrows exhibit the spin directions in the $S_z$-$S_x$ plane (top row) and the $S_y$-$S_x$ plane (bottom row).

On the other hand, we observed sample-dependent spin polarization signals across the five samples. As shown in Figs. 2(b)-(d), $S_x$ for the *x* = 5 %, 7.5 % and 10 % samples is opposite to that of the *x* = 2.5 % and 15 % samples. In the valence bands near the TSS gap, $S_y$ points out of the page at *k* > 0 and into the page at *k* < 0 for the *x* = 5 %, 7.5 % and 10 % samples, whereas the direction is reversed for the *x* = 2.5 % and 15 % samples. $S_z$, although primarily positive in the valence bands near the TSS gap for all samples, shows also sample-dependent polarizations farther from the gap, as it mainly points down (up) at *k* < 0 (*k* > 0) for *x* = 2.5 %, 10 % and 15 %, while pointing up (down) at *k* < 0 (*k* > 0) for *x* = 5 % and 7.5 %.

To further examine the distribution of the out-of-plane spin, distinguish between the helical-like and the hedgehog-like spin textures, and study the influence of photon energy and temperature on the spin configuration, we perform temperature-dependent SARPES measurements at $h\nu$ = 15 eV, with the spin direction set to $S_z$. The results are shown in Fig. 3. Fig. 3(a) presents the spin-integrated ARPES *k-E* maps and the corresponding second derivative analysis. From this panel, we can determine the doping levels of the samples based on the conduction band minimum and valence band maximum positions. Both rise from left to right, indicating increasing doping levels. By comparing the spin-resolved energy distribution curves (EDCs) at 10 K (below $T_N$) and 40 K (above $T_N$) in Fig. 3(b), we observe that the gap size is unchanged across $T_N$, consistent with our previous report[16], and that the sign of $S_z$ is also temperature-insensitive within the binding energy range we measured. This suggests that the out-of-plane spin polarization is the same for both the antiferromagnetic ground state and the paramagnetic state. By comparing the experimental results in Fig. 2 and Fig. 3(b), we find that at $h\nu$ = 6.3 eV, $S_z$ of both the conduction and valence bands is mainly pointing up. In contrast, at $h\nu$ = 15 eV, $S_z$ of the conduction and valence bands have opposite signs. The latter case is consistent with the reported hedgehog-like spin texture, indicative of the breaking of time-reversal symmetry and a magnetic origin of the TSS gap. We primarily attribute the universal spin-up out-of-plane spin polarization observed in our laser SARPES results to the final-state effects. Previous works reported that, when using circularly polarized 6 eV photons to probe the spin polarization of $Bi_2Se_3$, the final-state spin texture dominates the observed spin polarization, which changes according to the light polarization[27,29]. Although this phenomenon is slightly different from the situation in our experiment – the spin result shown in Fig. 2 is acquired using *p*-polarized light, whereas Ref. 29 used circularly polarized light – it is possible that *p*-polarized 6 eV photons produce similar final state effect, rendering the laser-SARPES result

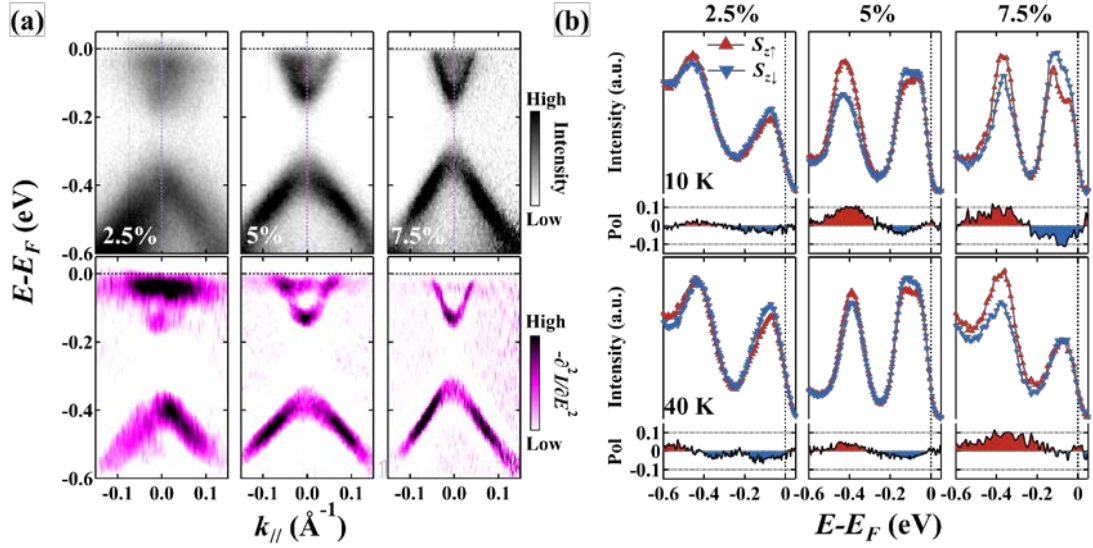

**Fig. 3. Spin-integrated and spin-resolved ARPES data acquired at $h\nu$ = 15 eV.** (a) Spin-integrated ARPES data measured using *p*-polarized light, and the corresponding second derivative analysis along the EDCs. The violet dash lines mark the measurement positions of the spin-resolved EDCs in (b). (b) $S_z$-resolved EDCs and polarization curves below and above the Néel temperature.

unrelated to the intrinsic spin configuration.

SARPES is the most direct way to study the spin texture of a material's energy bands. However, another technique, CD-ARPES, can be used to map out the configuration of the orbital angular momentum (OAM), and then infer the spin information of the material. Many studies have shown that in topological materials where spin and momentum are locked, such as Au[30-32], $Bi_2Se_3$[33–35], and $Bi_2Te_3$[36], CD-ARPES and SARPES results correspond well with each other[37–39]. Therefore, we also used CD-ARPES to examine the spin texture of Sb-doped $MnBi_2Te_4$, and the results are shown in Fig. 4. To clearly describe the CD pattern, we label the resolved conduction bands and valence bands from top to bottom as $C_3$, $C_2$, $C_1$, $V_1$, and $V_2$. Based on our previous work[16], $C_1$ and $V_1$ correspond to the TSS, and the other three bands correspond to the bulk states. For the four samples measured, the CD signals can be divided into three *k*-space regions: $R_1$ ($k < -0.04$ Å$^{-1}$), $R_2$ ($-0.04$ Å$^{-1}$ $< k < 0.04$ Å$^{-1}$) and $R_3$ ($k > 0.04$ Å$^{-1}$). Common characteristics include: (1) The TSSs above and below the gap have opposite CD signs. In $R_2$, the CD signs of $C_1$ and $V_1$ are opposite, consistent with the spin ARPES result in Fig. 3, supporting the hedgehog-like spin texture of the gapped TSS. This CD pattern differs from undoped $MnBi_2Te_4$, where the TSS shows an "X" shape and an antisymmetric CD sign about $\bar{\Gamma}$[40,41], indicating that Sb doping alters the OAM near the Dirac point. (2) In $R_1$ and $R_3$, the CD signs of $C_1$, $V_1$ and $V_2$ are opposite, consistent with the helical spin texture found in Fig. 2. (3) Temperature has little effect on the CD pattern. Comparing CD-ARPES results below and above $T_N$ [Fig. 4(b) and 4(c)], the CD pattern remains unchanged, in line with previous results for undoped $MnBi_2Te_4$[41]. Sample-dependent signals include: (1) the valence bands show three types of patterns: for the $x = 7.5\%\_1$ sample, the pattern is "red-blue-blue" from $R_1$ to $R_3$; for the $x = 7.5\%\_2$ and 10% samples, it is "red-red-blue"; for $x = 15\%$, it is "blue-red-red". The conduction band $C_2$ also shows varying CD patterns: for $x = 7.5\%\_1$, it is mostly red; for $x = 7.5\%\_2$, it alternates between red (for $k < 0$) and blue (for $k > 0$); for $x = 10\ \%$, this pattern is opposite to that of $x =$ 7.5%_2. The CD pattern for the conduction band C$_3$ is detected only for $x$ = 7.5%_1 and 7.5%_2. For the former, C$_3$ is entirely blue; for the latter, C$_3$ shows a left-blue, right-red pattern. (2) High doping concentrations may invert the OAM direction. Comparing the CD signs in R$_1$ and R$_3$, the CD sign of $x$ = 15% sample is opposite to that of other samples, indicating that its OAM has inverted. However, the sign of the TSSs in the R$_2$ region is the same for $x$ = 7.5%_2, 10% and 15%, suggesting that doping has little effect on the OAM of TSSs near the Dirac point.

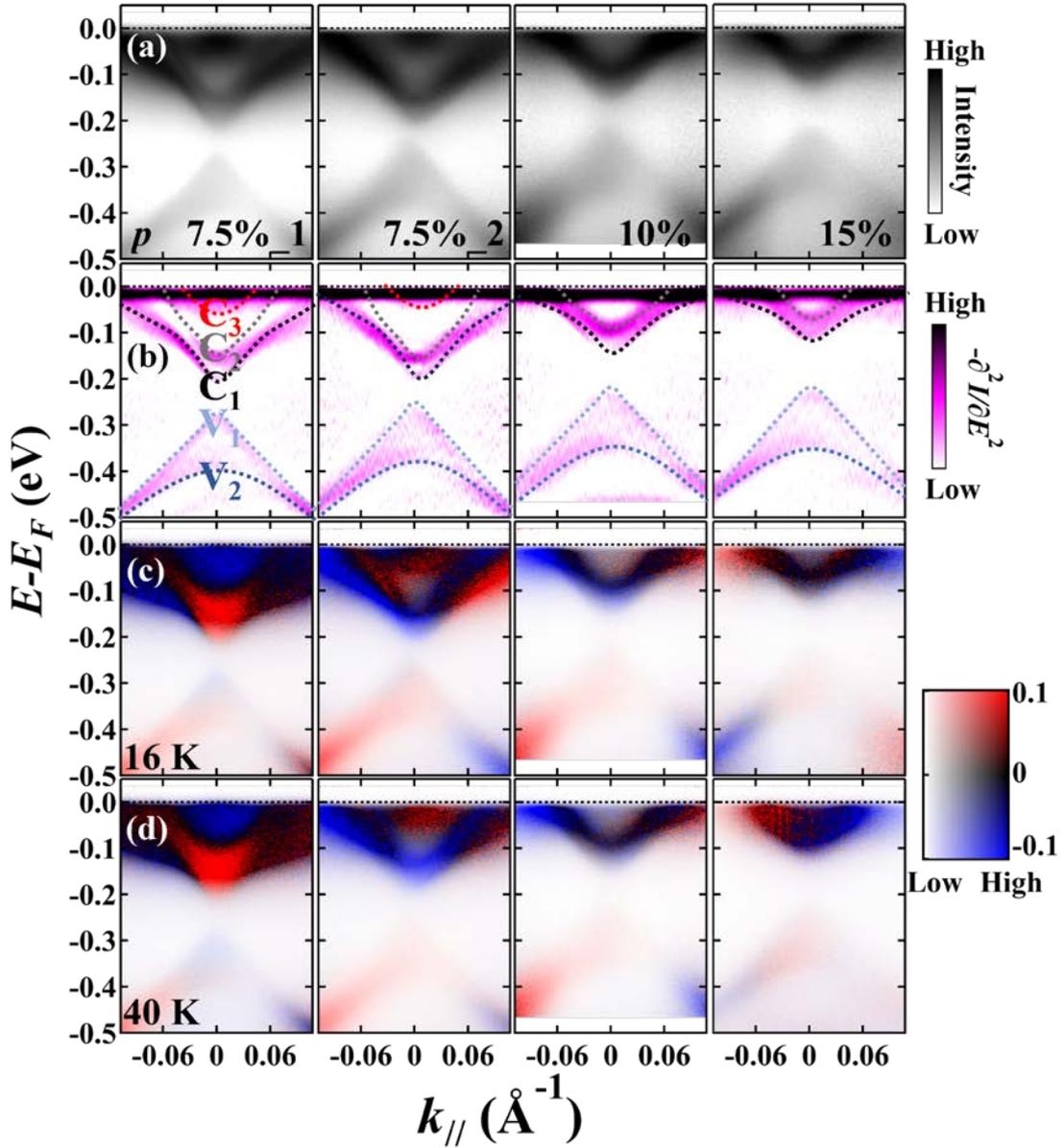

**Fig. 4. CD-ARPES data for Mn(Bi$_{1-x}$Sb$_x$)$_2$Te$_4$ with different $x$ and temperatures, acquired at $h\nu$ = 6.3 eV.** (a) Spin-integrated ARPES data measured using $p$-polarized laser radiation. (b) Corresponding second derivative analysis along the EDCs. C$_1$-C$_3$ mark the three conduction bands and V$_1$, V$_2$ mark the two valence bands. (c-d) CD-ARPES maps below and above the Néel temperature. The intensity of the band structure collected by left- and right-circularly polarized light is defined as $I_L$ and $I_R$, respectively. The CD intensity is defined as $I_{CD} = (I_L - I_R) / (I_L + I_R)$, where different signs of $I_{CD}$ represent different OAM states of the band structure.

Now we discuss the possible reasons for this sample-dependent hedgehog-like spin configuration. Three key phenomena need to be elaborated. First, we discuss the origination of the hedgehog-like spin configuration in this system. Such spin texture is previously reported on Mn-doped $Bi_2Se_3$ thin films[18]. In that system, magnetic doping breaks the time-reversal symmetry of the surface, lifting the Kramers degeneracy, resulting in a nonzero out-of-plane spin polarization at $\bar{\Gamma}$. A similar mechanism may also explain the spin configuration in Sb-doped $MnBi_2Te_4$. Numerous experiments have demonstrated that $MnBi_2Te_4$ possesses helical Rashba-type in-plane spin texture[42–44]. Sb substitution may modify the magnetic ground state of $MnBi_2Te_4$, breaking the time-reversal symmetry of the surface, which leads to the opposite spin polarization observed in Fig. 3(c). By careful band analysis, we attribute the opposite $S_z$ component near the band gap mainly to the TSS. As shown in the spin-EDCs of Fig. 3(b), the conduction band displays two peaks. Comparing these with the $k$-$E$ map in Fig. 3(a), we found that the peak near the Fermi level corresponds to the bulk band $C_2$, and the peak near the Dirac point corresponds to the surface state $C_1$. In the valence band, only one peak is resolved, corresponding to the surface state $V_1$. At $h\nu = 15$ eV, the $S_z$ component of $C_1$ is negative, while that of $V_1$ is positive, consistent with the hedgehog-like spin texture. This is also supported by the CD results in Fig. 4, in which the CD signals of the surface state have opposite signs at the energies near the Dirac point. Far from the Dirac point, the CD signal exhibits a helical OAM texture, in line with observations in Mn-doped $Bi_2Se_3$[18]. In light with the facts that SARPES with higher incident photon energies tend to exhibit intrinsic spin configuration of the samples, and that our CD-ARPES data are in line with the SARPES results at $h\nu = 15$ eV, we believe that the spin configuration of Sb-doped $MnBi_2Te_4$ is consistent with the hedgehog-like spin texture.

Regarding the sample dependence of the spin helicity, we hypothesize that we may have observed the hidden spin polarization (HSP) in this system. The so-called HSP is a theoretical model describing the observed spin polarization in centrosymmetric crystals induced by the breaking of local inversion symmetry[45]. Recently, this theory is extended to antiferromagnets[46]. Through symmetry analysis, we found that the symmetry of Sb-doped $MnBi_2Te_4$ meets HSP requirements. Its bulk space group is centrosymmetric but the site point group of Bi and Te ($C_{3v}$) is non-centrosymmetric, the combination of which leads to the HSP where the unit cell is spatially separated into two sectors carrying opposite spin polarizations[45]. Because the laser's spot size is very small (less than 10 μm[47]), there is a high possibility that the cleaved surfaces within the probe area originate from a single sector. The cleaved surface varies across different samples, leading to the detection of different sectors and, consequently, the observation of different spin polarizations. This explanation is consistent with our experimental results.

In conclusion, we examine the spin configuration of Sb-doped $MnBi_2Te_4$ by both SARPES and CD-ARPES. We found sample-dependent hedgehog-like spin configuration in this system, consistent with the existence of time-reversal symmetry breaking and the magnetic origin of the TSS gap. Our results imply the possibility of manipulating the spin states and realizing high-temperature QAHE in Sb-doped $MnBi_2Te_4$.

*Experimental Methods*

*Single Crystal Growth and XRD Characterization.* High-quality single crystals of Sb-doped MnBi$_2$Te$_4$ were grown using the self-flux method[16]. The samples were characterized by XRD at room temperature using a Rigaku SmartLab diffractometer with Cu K$\alpha$ radiation. The diffraction pattern in Fig. 1(b) confirms that the cleaved surfaces of the crystal are parallel to the (000*l*) crystallographic orientation.

*ARPES measurements.* SARPES experiments were performed at Hiroshima Synchrotron Radiation Facilities (HiSOR), Hiroshima, Japan, using the LaserSpin ARPES endstation[47] with a Scienta DA30 analyzer, and the ESPRESSO endstation at BL9B[48] with a VG-SCIENTA R4000 analyzer. A laser light with photon energy $h\nu$ = 6.3 eV was used in the former, while synchrotron lights with photon energies $h\nu$ = 15 eV and 28 eV were used in the latter. CD-ARPES experiments were performed at HiSOR using the $\mu$-ARPES[49] endstation with a VG-SCIENTA R4000 analyzer and a laser with $h\nu$ = 6.3 eV. All the samples were cleaved *in-situ* along the (000*l*) crystal plane in an ultrahigh vacuum better than 7×10$^{-11}$ mbar.

*Acknowledgments.* We thank Yudai Miyai, Amit Kumar, and Yogendra Kumar for the help in $\mu$-ARPES measurements in Hiroshima Synchrotron Radiation Center (HISOR), Hiroshima, Japan. Work at SUSTech was supported by the National Key R&D Program of China (no. 2022YFA1403700), the National Natural Science Foundation of China (nos. 12074161 and 12204221), and the Guangdong Basic and Applied Basic Research Foundation (No. 2022A1515012283). C.L. acknowledges the Highlight Project (No. PHYS-HL-2020-1) of the College of Science at SUSTech. The ARPES experiments were performed at BL9B in HiSOR under the approval of Proposal No. 22AG037, and at BL13U of the National Synchrotron Radiation Laboratory (NSRC), Hefei, China under the approval of Proposal 2021-HLS-PT-003656.

# References

1. C. Z. Chang, J. Zhang, X. Feng, *et al*. Experimental observation of the quantum anomalous Hall effect in a magnetic topological insulator. *Science* **2013**, 340(6129): 167-170.
2. X. L. Qi, T. L. Hughes, S. C. Zhang. Topological field theory of time-reversal invariant insulators. *Phys. Rev. B* **2008**, 78(19): 195424.
3. R. Yu, W. Zhang, H. J. Zhang, *et al*. Quantized anomalous Hall effect in magnetic topological insulators. *Science* **2010**, 329(5987): 61-64.
4. C. Liu, Y. Wang, H. Li, *et al*. Robust axion insulator and Chern insulator phases in a two-dimensional antiferromagnetic topological insulator. *Nat. Mater.* **2020**, 19(5): 522-527.
5. N. P. Armitage, L. Wu. On the matter of topological insulators as magnetoelectrics. *SciPost Phys.* **2019**, 6(4): 046.
6. L. Wu, M. Salehi, N. Koirala, *et al*. Quantized Faraday and Kerr rotation and axion

electrodynamics of a 3D topological insulator. *Science* **2016**, 354(6316): 1124-1127.

7. X. L. Qi, S. C. Zhang. The quantum spin Hall effect and topological insulators. *Phys. Today* **2010**, 63(1): 33-38.
8. D. Steffensen, M. H. Christensen, B. M. Andersen, *et al*. Topological superconductivity induced by magnetic texture crystals. *Phys. Rev. Res.* **2022**, 4(1): 013225.
9. K. Nomura, N. Nagaosa. Surface-quantized anomalous Hall current and the magnetoelectric effect in magnetically disordered topological insulators. *Phys. Rev. Lett.* **2011**, 106(16): 166802.
10. M. M. Otrokov, I. I. Klimovskikh, H. Bentmann, *et al*. Prediction and observation of an antiferromagnetic topological insulator. *Nature* **2019**, 576(7787): 416-422.
11. Y.-J. Hao, P. Liu, Y. Feng, *et al*. Gapless surface Dirac cone in antiferromagnetic topological insulator $MnBi_2Te_4$. *Phys. Rev. X* **2019**, 9(4): 041038.
12. P. Swatek, Y. Wu, L. L. Wang, *et al*. Gapless Dirac surface states in the antiferromagnetic topological insulator $MnBi_2Te_4$. *Phys. Rev. B* **2020**, 101(16): 161109.
13. Y. J. Chen, L. X. Xu, J. H. Li, *et al*. Topological electronic structure and its temperature evolution in antiferromagnetic topological insulator $MnBi_2Te_4$. *Phys. Rev. X* **2019**, 9(4): 041040.
14. H. Li, S. Y. Gao, S. F. Duan, *et al*. Dirac surface states in intrinsic magnetic topological insulators $EuSn_2As_2$ and $MnBi_{2n}Te_{3n+1}$. *Phys. Rev. X* **2019**, 9(4): 041039.
15. Y. Deng, Y. Yu, M. Z. Shi, *et al.* Quantum anomalous Hall effect in intrinsic magnetic topological insulator $MnBi_2Te_4$. *Science* **2020**, 367(6480): 895-900.
16. X.-M. Ma, Y. Zhao, K. Zhang, *et al*. Realization of a tunable surface Dirac gap in Sb-doped $MnBi_2Te_4$. *Phys. Rev. B* **2021**, 103(12): L121112.
17. L. Pan, X. Liu, Q. L. He, *et al.* Probing the low-temperature limit of the quantum anomalous Hall effect. *Sci. Adv.* **2020**, 6(25): eaaz3595.
18. S.-Y. Xu, M. Neupane, C. Liu, *et al*. Hedgehog spin texture and Berry's phase tuning in a magnetic topological insulator. *Nat. Phys*. **2012**, 8(8): 616-622.
19. B. Chen, F. Fei, D. Zhang, *et al*. Intrinsic magnetic topological insulator phases in the Sb doped $MnBi_2Te_4$ bulks and thin flakes. *Nat. Commun.* **2019**, 10(1): 4469.
20. J. Q. Yan, S. Okamoto, M. A. McGuire, *et al*. Evolution of structural, magnetic, and transport properties in $MnBi_{2-x}Sb_xTe_4$. *Phys. Rev. B* **2019**, 100(10): 104409.
21. M. H. Du, J. Yan, V. R. Cooper, *et al.* Tuning Fermi levels in intrinsic antiferromagnetic topological insulators $MnBi_2Te_4$ and $MnBi_4Te_7$ by defect engineering and chemical doping. *Adv. Funct. Mater.* **2021**, 31(3): 2006516.
22. S. Mukherjee, A. K. NM, S. Manna, *et al*. Magnetic order influenced phonon and electron dynamics in $MnBi_2Te_4$ and Sb-doped $MnBi_2Te_4$ investigated by terahertz time-domain spectroscopy. *Phys. Rev. B* **2024**, 110(19): 195401.
23. D. A. Glazkova, D. A. Estyunin, I. I. Klimovskikh, *et al*. Electronic structure of magnetic topological insulators $Mn(Bi_{1-x}Sb_x)_2Te_4$ with various concentration of Sb atoms. *JETP Lett.* **2022**, 115(5): 286-291.
24. H. H. Wang, X. G. Luo, M. Z. Shi, *et al*. Possible bipolar effect inducing anomalous transport behavior in the magnetic topological insulator $Mn(Bi_{1-x}Sb_x)_2Te_4$. *Phys. Rev. B* **2021**, 103(8): 085126.
25. D. A. Glazkova, D. A. Estyunin, I. I. Klimovskikh, *et al*. Mixed type of the magnetic order in intrinsic magnetic topological insulators $Mn(Bi,Sb)_2Te_4$. *JETP Lett.* **2022**, 116(11): 817-824.
26. H. Iwasawa, K. Shimada, E. F. Schwier, *et al*. Rotatable high-resolution ARPES system for

tunable linear-polarization geometry. *J. Synchrotron Radiat.* **2017**, 24(4): 836-841.
27. C. Jozwiak, C. H. Park, K. Gotlieb, *et al*. Photoelectron spin-flipping and texture manipulation in a topological insulator. *Nat. Phys.* **2013**, 9(5): 293-298.
28. A. M. Shikin, D. A. Estyunin, I. I. Klimovskikh, *et al*. Nature of the Dirac gap modulation and surface magnetic interaction in axion antiferromagnetic topological insulator $MnBi_2Te_4$. *Sci. Rep.* **2020**, 10(1): 13226.
29. J. Sánchez-Barriga, A. Varykhalov, J. Braun, *et al*. Photoemission of $Bi_2Se_3$ with circularly polarized light: Probe of spin polarization or means for spin manipulation? *Phys. Rev. X* **2014**, 4(1): 011046.
30. S. R. Park, C. H. Kim, J. Yu, *et al*. Orbital-angular-momentum based origin of Rashba-type surface band splitting. *Phys. Rev. Lett.* **2011**, 107(15): 156803.
31. H. Ryu, I. Song, B. Kim, *et al*. Photon energy dependent circular dichroism in angle-resolved photoemission from Au(111) surface states. *Phys. Rev. B* **2017**, 95(11): 115144.
32. M. Nagano, A. Kodama, T. Shishidou, *et al*. A first-principles study on the Rashba effect in surface systems. *J. Phys. : Condens. Matter* **2009**, 21(6): 064239.
33. Y. H. Wang, D. Hsieh, D. Pilon, *et al*. Observation of a warped helical spin texture in $Bi_2Se_3$ from circular dichroism angle-resolved photoemission spectroscopy. *Phys. Rev. Lett.* **2011**, 107(20): 207602.
34. J. Zhang, J. Caillaux, Z. Chen, *et al*. Probing spin chirality of photoexcited topological insulators with circular dichroism: multi-dimensional time-resolved ARPES on $Bi_2Te_2Se$ and $Bi_2Se_3$. *J. Electron Spectrosc. Relat. Phenom.* **2021**, 253: 147125.
35. F. Vidal, M. Eddrief, B. Rache Salles, *et al*. Photon energy dependence of circular dichroism in angle-resolved photoemission spectroscopy of $Bi_2Se_3$ Dirac states. *Phys. Rev. B* **2013**, 88(24): 241410.
36. Y. Wang, N. Gedik. Circular dichroism in angle-resolved photoemission spectroscopy of topological insulators. *Phys. Status Solidi RRL* **2013**, 7(1-2): 64-71.
37. M Schüler, U. De Giovannini, H. Hübener, *et al*. Local Berry curvature signatures in dichroic angle-resolved photoelectron spectroscopy from two-dimensional materials. *Sci. Adv.* **2020**, 6(9): eaay2730.
38. S. Cho, J. H. Park, J. Hong, *et al*. Experimental observation of hidden berry curvature in inversion-symmetric bulk $2H-WSe_2$. *Phys. Rev. Lett.* **2018**, 121(18): 186401.
39. G. Schönhense. Circular dichroism and spin polarization in photoemission from adsorbates and non-magnetic solids. *Phys. Scr.* **1990**, 1990(T31): 255.
40. A. Liang, C. Chen, H. Zheng, *et al*. Approaching a minimal topological electronic structure in antiferromagnetic topological insulator $MnBi_2Te_4$ via surface modification. *Nano Lett.* **2022**, 22(11): 4307-4314.
41. C. Yan, S. Fernandez-Mulligan, R. Mei, *et al*. Origins of electronic bands in the antiferromagnetic topological insulator $MnBi_2Te_4$. *Phys. Rev. B* **2021**, 104(4): L041102.
42. R. C. Vidal, H. Bentmann, T. R. F. Peixoto, *et al*. Surface states and Rashba-type spin polarization in antiferromagnetic $MnBi_2Te_4$ (0001). *Phys. Rev. B* **2019**, 100(12): 121104.
43. N. L. Zaitsev, I. P. Rusinov, T. V. Menshchikova, *et al*. Interplay between exchange-split Dirac and Rashba-type surface states at the $MnBi_2Te_4$/BiTeI interface. *Phys. Rev. B* **2023**, 107(4): 045402.
44. R. Xu, Y. Bai, J. Zhou, *et al*. Evolution of the electronic structure of ultrathin $MnBi_2Te_4$ films.

*Nano Lett.* **2022**, 22(15): 6320-6327.

45. X. Zhang, Q. Liu, J. W. Luo, *et al*. Hidden spin polarization in inversion-symmetric bulk crystals. *Nat. Phys.* **2014**, 10(5): 387-393.
46. L. D. Yuan, X. Zhang, C. M. Acosta, *et al*. Uncovering spin-orbit coupling-independent hidden spin polarization of energy bands in antiferromagnets. *Nat. Commun.* **2023**, 14(1): 5301.
47. T. Iwata, T. Kousa, Y. Nishioka, *et al*. Laser-based angle-resolved photoemission spectroscopy with micrometer spatial resolution and detection of three-dimensional spin vector. *Sci. Rep.* **2024**, 14(1): 127.
48. T. Okuda, K. Miyamoto, A. Kimura, *et al*. A double VLEED spin detector for high-resolution three-dimensional spin vectorial analysis of anisotropic Rashba spin splitting. *J. Electron Spectrosc. Relat. Phenom.* **2015**, 201: 23-29.
49. H. Iwasawa, E. F. Schwier, M. Arita, *et al*. Development of laser-based scanning *μ*-ARPES system with ultimate energy and momentum resolutions. *Ultramicroscopy* **2017**, 182: 85-91.